\pdfoutput=1
\documentclass[%
reprint, notitlepage,
superscriptaddress, onecolumn,
 amsmath,amssymb,
 aps,
]{revtex4-2}

\usepackage{graphicx}  
\usepackage[bookmarks=true, hidelinks, breaklinks=true]{hyperref}
\usepackage{xcolor}
\usepackage[normalem]{ulem}  
\usepackage{breakcites} 
\usepackage{bm}
\usepackage[utf8]{inputenc}
\usepackage{etoolbox}
\usepackage{textcomp}
\usepackage{caption}

\begin{document}

\title{Rupture dynamics of flat colloidal films}

\author{Phalguni Shah}
\affiliation{Department of Physics and Astronomy, Northwestern University, Evanston, IL, United States}

\author{Eleanor Ward}
\affiliation{Department of Physics and Astronomy, Northwestern University, Evanston, IL, United States}
\affiliation{Currently at Department of Mathematics, Georgetown University}

\author{Srishti Arora}
\affiliation{Department of Physics and Astronomy, Northwestern University, Evanston, IL, United States}
\affiliation{Currently at Institute for New Materials, Saarbrücken, Germany}

\author{Michelle M.\ Driscoll}
\email[Corresponding Author:]{michelle.driscoll@northwestern.edu}
\affiliation{Department of Physics and Astronomy, Northwestern University, Evanston, IL, United States}

\begin{abstract}
Here, we report experimental results on the rupture of flat colloidal films over a large range of volume fractions, 0.00 $\le \phi\le$ 0.47. The films are formed using a constant fluid volume, ruptured with a needle, and recorded using a high-speed camera. We show that colloidal films rupture in a manner quantitatively similar to Newtonian fluids, even well into the shear thinning regime. These results are consistent with the well-known mechanism of the rupture of Newtonian films, where the rupture rim rolls outward collecting more fluid and thus film rupture is a shear-free process. However, in the case of spontaneous rupture under controlled humidity conditions, the same dense colloidal films show exotic instabilities reminiscent of a wrinkling fabric on the film surface. These instabilities were absent in manually ruptured films. We hypothesize that these instabilities occur when the film thickness becomes thin enough to compete with the colloidal particle size, due to film drainage before spontaneous rupture. Thus, although non-Newtonian flow properties do not influence film rupture dynamics for thick enough films, the effect of microstructure has dramatic consequences in thinner films.
\end{abstract}
\maketitle

\section{Introduction}

Foams, bubbles and films are ubiquitous in processes that involve liquid-gas interfaces, ranging from detergents~\cite{kissa2020detergency}, volcanic eruptions~\cite{Cassidy2018volcano}, and water-borne disease transmission~\cite{BourouibaDisease}. Fluid films are also a convenient system to study the dynamics of fluids in a two-dimensional geometry. Newtonian bubbles have been studied in detail for several decades~\cite{mysels1959soap, mcentee1969bursting, frankel1969bursting, taylor1959dynamics,culick1960comments,Debregeas1704, DebregeasPRL}, and their rupture dynamics in the inviscid limit are well understood. For low-viscosity Newtonian soap films of constant thickness, Culick~\cite{culick1960comments} derived the rupture speed of an inviscid film as:

\begin{equation}\label{eq:culick}
    u_c=\sqrt{\frac{2\sigma}{\rho h}}.
\end{equation}
This is commonly known as the Culick's law, and it has been experimentally verified for a  variety of Newtonian fluid films~\cite{mcentee1969bursting}. In contrast to the linear rupture growth predicted by Culick for low-viscosity films, experiments on higher viscosity liquid curtains have observed a slowing down in rupture growth~\cite{LiquidCurtainVisocous}. Moreover, the rupture of extremely viscous bubbles (a million times more viscous than water) has been observed to grow exponentially~\cite{DebregeasPRL,Debregeas1704}. Initially, viscoelastic effects were suggested as a cause of this exponential growth, but consequent theoretical work showed that high-viscosity films grow exponentially over a transient timescale directly proportional to fluid viscosity, before asymptotically achieving the Culick velocity~\cite{savva2009viscous}. This prediction is thus quantitatively consistent with the experimental data for extremely viscous films~\cite{DebregeasPRL,Debregeas1704}.

A majority of processes involve bursting films that contain surfactants, and/or made of non-Newtonian fluids; surfactants and particulate additives can significantly alter rupture dynamics.  
For example, films with high surfactant concentrations (above the critical micellar concentration) rupture slower than the Culick velocity, and can develop ridges, mesas~\cite{zhangSharma2015domain,VivekSharma_mesa_2021}, or crack-like instabilities~\cite{petit2015holes}. Films and bubbles composed of smectic materials can even develop reversible instabilities under stress ~\cite{muller2007smecticBurst, Shape_smectic_2022} and viscoelastic films display flowering instabilities at the rupture rim ~\cite{flowering_viscoelastic_2021}. In the ultrathin limit ($\sim10$ nm), even Newtonian soap films exhibit viscoelastic properties that have been attributed to the competing lengthscale of film thickness and the size of surfactant molecules~\cite{evers_NewtonBlack_elastic_1997}. It is also known that films laden with particles larger than film thickness rupture intermittently due to the presence of particles~\cite{timounay2015particulate}. Thus, both micelles and particulate additives alter film dynamics significantly,  and this effect is more dramatic in the presence of the competing lengthscales of film thickness and additive size. 

Here, we report experimental data on the rupture of colloidal soap films over a large range of colloidal volume fraction: $0\le\phi\le0.47$. Our experiments enable us to access phenomena caused by bulk fluid properties, as well as more striking behaviors due to discrete effects introduced on the scale of particle size. We find that manually ruptured films behave like Newtonian viscous films and rupture at a constant speed, while thinner and spontaneously rupturing films develop exotic instabilities and rupture in a qualitatively different fashion. Similar to past studies, these instabilities may be a result of the particle size being comparable to the film thickness.

\begin{figure}[!thbp]
\centering
\includegraphics[width=\textwidth
]{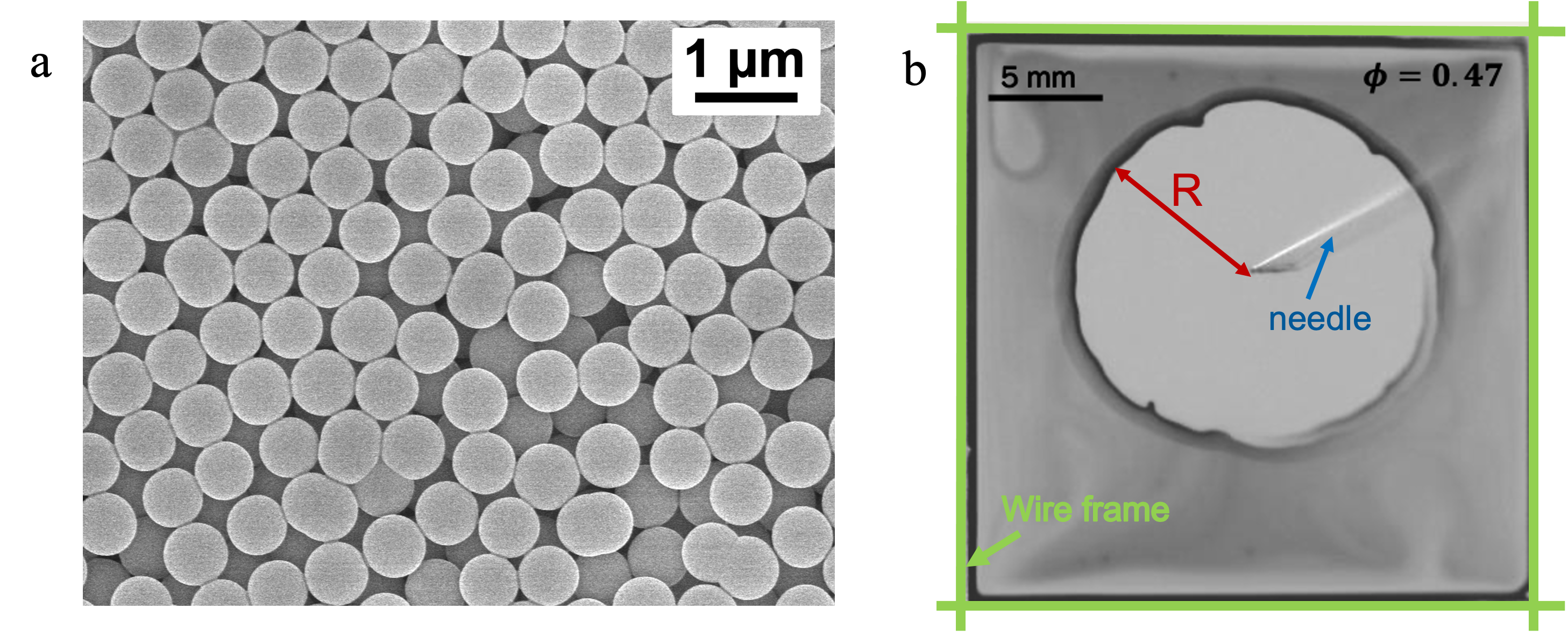}
    \caption{(a) SEM image of the colloidal silica spheres. The spheres were suspended in the mixture of SDS and water to make suspensions with volume fractions $0\le\phi\le0.47$, and the fluids were used to form horizontal films and record their rupture. (b) A background-divided snapshot of a $\phi=0.47$ film during rupture. After being manually ruptured with a needle, a circular hole forms and its radius, R, grows until the boundary interferes.  
}
    \label{fig:pictures}
\end{figure}

For manually ruptured films, the rupture data follows a similar trend as Newtonian films of similar viscosity when plotted in terms of the effective viscosity of the colloidal fluid. Our results demonstrate that rupture dynamics of colloidal films can be characterized by effective Newtonian viscosity, and additionally that viscosity has a significant role in the thickness profile of constant-volume films. Furthermore, we report exotic instabilities in spontaneously rupturing dense colloidal films. These instabilities occur in environments of both controlled and uncontrolled humidity, but they are consistently reproducible only in a humidity-controlled environment. We hypothesize that these unstable structures develop when the film thickness becomes comparable to the colloidal size. Thus, this system may help us uncover the transition from continuum dynamics to discrete effects in colloidal fluids. 

\section{Experimental Methods}

\noindent \textbf{Colloidal synthesis}\\

The silica colloidal spheres were synthesized following the St\"ober process~\cite{stober1968,zhang2009}. To initiate the reaction, we mixed TEOS (tetraethyl orthosilicate) with ethanol and water at room temperature and in presence of ammonia as a catalyst. Subsequently, we added a feed of TEOS and water to increase the particle size. We used the total number of feeds as the means to control particle diameter. The colloids were then washed with ethanol, separated to decrease polydispersity, and re-suspended in water. The average silica sphere diameter used for experiments reported here was 660 $\pm$ 20 nm [Fig. \ref{fig:pictures}a]. Suspensions of volume fractions $0.00\le\phi\le0.47$ were prepared in water containing 4 mM (a concentration well below the critical micellar concentration of 8 mM) Sodium Dodecyl Sulfate (SDS), a surfactant. Thus, we do not expect SDS to affect the system in any way other than decreasing the overall surface tension. To compare colloidal rupture data to Newtonian fluid films, we perform rupture experiments on water-glycerol mixtures in the presence of 4 mM SDS, so that the fluid viscosity ranges from 1 cP to 235 cP~\cite{sheely1932glycerol}.\\

\begin{figure}[!hbt]
\centering
\includegraphics[width=\textwidth
]{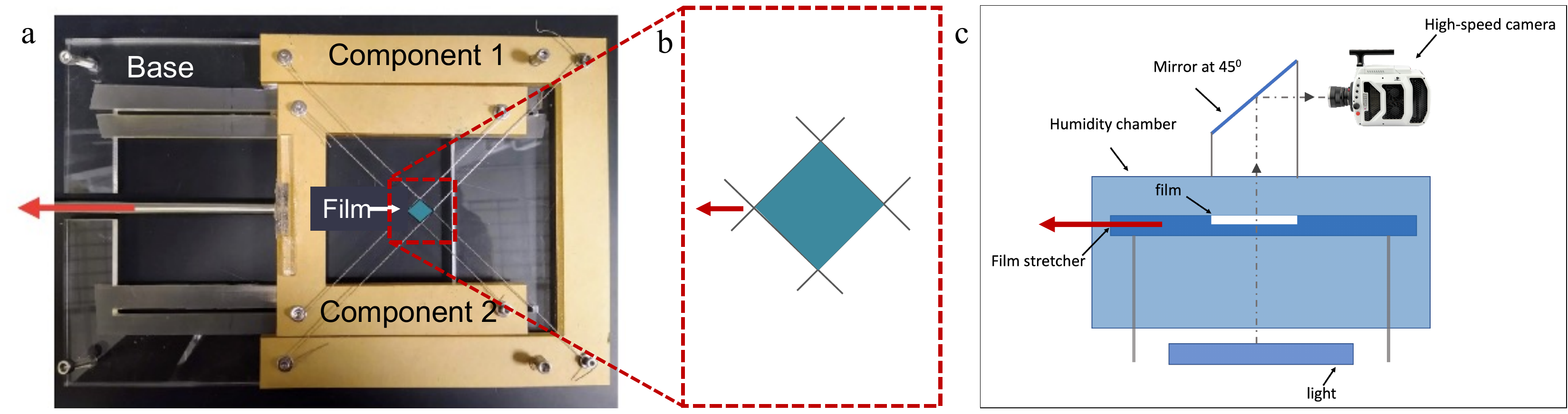}
    \caption [Experimental setup for colloidal film rupture.]{(a) Top view of the film stretcher designed to make films of a desired size and reproducible thickness (see also SI video 1~\cite{filmSI}). (b) A known volume of fluid is introduced between the crosswires, and then the stretcher is drawn using a motor to expand the crosswires and form a film of desired size. We used the fluid volumes of $10 \mu l$ and $15 \mu l$ in our experiments. (c) Side view of the experimental setup. The film stretcher is enclosed in a chamber that maintains relative humidity between $ 75\%$ and $85\%$ during experiments. The horizontal film is lit from below and its rupture is filmed using a high-speed camera. 
}
    \label{fig:setup}
\end{figure}

\noindent \textbf{Experimental setup}\\

Experiments on film rupture commonly involve spherical bubbles formed from a reservoir~\cite{Poulain2018Biosurfactants, DebregeasPRL,Debregeas1704} or vertical films~\cite{mysels1959soap, mcentee1969bursting}. However, as evaporation effects are more pronounced for dense colloidal suspensions, the reservoir setup is not convenient for colloidal films. Therefore, in our experiments, we form flat horizontal films by introducing a constant volume of fluid onto a custom film stretcher inspired by past experiments with bacterial films~\cite{sokolov2010swimming}.

To form horizontal films of reproducible and controlled thickness, we built a custom film stretcher inspired from Sokolov et al~\cite{sokolov2010swimming}(see also SI video 1~\cite{filmSI}). The stretcher is made of acrylic pieces cut using a laser cutter. On the stretcher base sit two U-shaped components, so that Component 2 fits inside Component 1 [Fig. \ref{fig:setup}a]. Each component has taut metal wires that criss-cross to form a small square hole. Component 1 is stationary, while Component 2 is attached to a motor and can be moved horizontally at a constant speed. A known volume of fluid can be introduced on the small square hole formed by the wires. When component 2 is pulled by a motor, the square hole expands, allowing us to form a horizontal square film of desired size. [Fig. \ref{fig:setup}b] For all experiments reported here, Component 2 was pulled at the speed of of 0.8 mm/s. To minimize the effects of air currents, impurities, and evaporation, the film stretcher was mounted inside a custom humidity chamber and the relative humidity was maintained between 75\% and 85\% using a reservoir of NaCl solution inside the chamber [Fig. \ref{fig:setup}c]. The film size was 25 mm $\times$ 25 mm, and films were formed using two fluid volumes: $10$ \textmu L and $15$ \textmu L. The horizontal films were illuminated from below using a white panel light and the transmitted light data was recorded at 83,000 fps using a Phantom v2512 high-speed camera. The rupture was manually initiated near the center of the film using a needle. For the viscous film rupture data, we made solutions with known concentrations of glycerol and water by mass, and used a constant SDS concentration of 4 mM in all the mixtures. These Newtonian films were also formed using the same film stretcher, and with fluid volumes of 10 \textmu L and 15 \textmu L.\par

\noindent \textbf{Film profile characterization}

In order to compare film thickness profiles for films of different fluid viscosities, we used two different techniques: interference imaging and dye absorbance. For the interference imaging, a green filter of wavelength $530 \pm 10$ nm was introduced in the path of the light, and films of two different viscosities were imaged before rupture was initiated. For the dye absorption measurements,  we dissolved 10 g/L of Brilliant Blue dye (Erioglaucine  disodium salt) in the mixtures that contained varying amounts of glycerol and water, in presence of 4 mM SDS. We imaged these dyed films under transmitted white light, using the same high-speed camera as the one used to collect the rupture velocity data. 

\noindent \textbf{Data analysis}

Image processing and analysis were performed using ImageJ, and data was plotted using python. To measure the rupture velocity, the frame of rupture initiation was identified, and the distance from the initiation point to the rupture rim was measured for every 10$^{th}$ frame, until boundary conditions affect the rupture, approximately when rupture radius is around 10 mm (for the wire frame size of 25 mm). 
The dye absorbance images were background-normalized with the last frame of the image sequence, where the film was absent due to completion of rupture. A rectangular region, 10 pixels in height and the length of the film in width, was selected near the film center, and the line profile, averaged over the 10 pixels, was plotted. This procedure was repeated for different viscosity films.

\section{Results and discussion}
\subsection{Rupture velocity: colloidal vs. viscous Newtonian films}

\begin{figure}[!htb]
\centering
\includegraphics[width=\textwidth
]{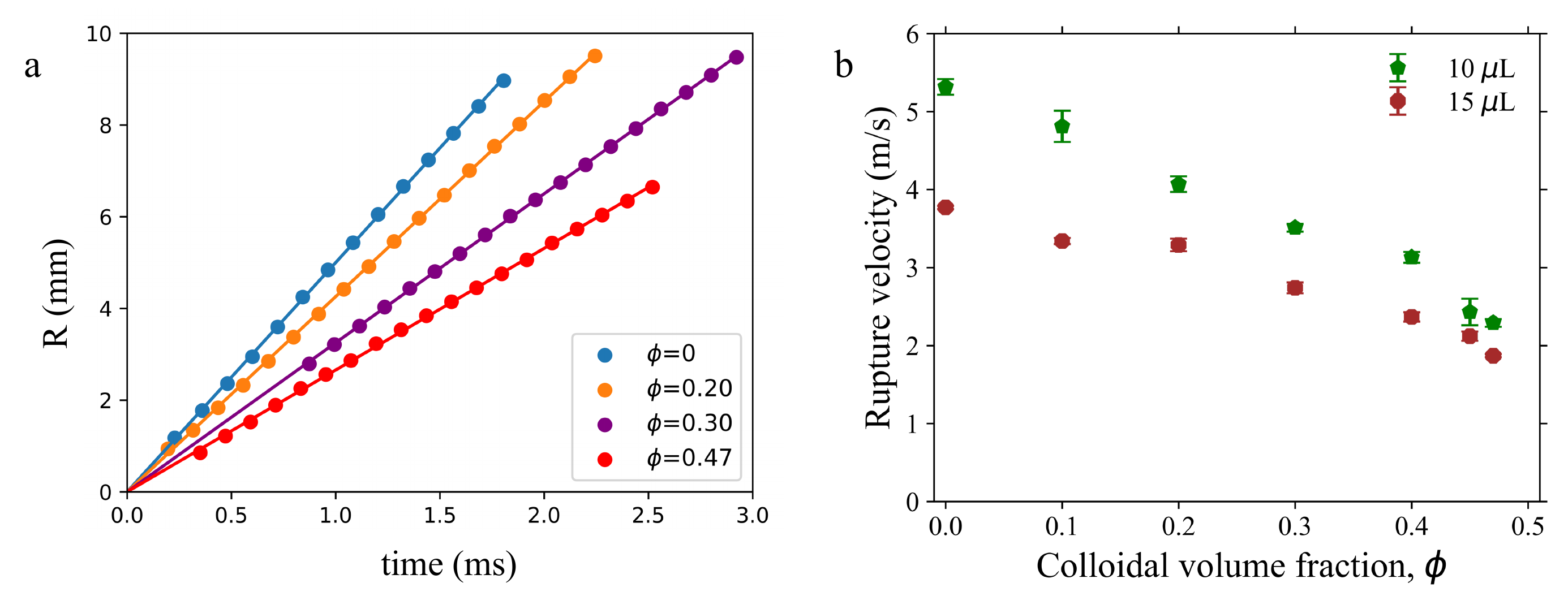}
    \caption[Rupture velocity of colloidal soap films for $0.0\le\phi\le0.47$.] {(a) Rupture radius vs time for $15 \mu l$ films made of fluids with increasing  colloidal volume fraction $\phi$.  The rupture velocity (slope of the linear fit lines) is surprisingly constant even at $\phi$  where the fluid is highly non-Newtonian. (b) The film rupture velocity, plotted against $\phi$ for two fluid volumes, decreases with increasing $\phi$. Error bars are standard deviations over at least 5 trials.  }
    \label{fig:colloid_rupturedata}
\end{figure}

Figure \ref{fig:pictures}b shows a snapshot of a colloidal film mid-rupture. When the rupture is induced in the center of the film, a circular hole forms in the film and grows in size. We measure the radius of this hole, $R$, with respect to time, $t$, for suspension films of increasing $\phi$. In Figure \ref{fig:colloid_rupturedata}a, we report this data, truncated at the point in time when the effect of film boundary destroys the circular symmetry of the rupture (when the rupture is about 80\% the size of the film). Surprisingly, even at very high $\phi$ where the bulk colloidal suspension is known to be highly non-Newtonian, the $R$ vs.\ $t$ data follows a linear trend indicating a constant rupture velocity over time. The lines in Figure \ref{fig:colloid_rupturedata}a indicate linear fits to the $R$ vs. $t$ data, and their slope thus gives the rupture velocity values.  Though it remains constant throughout rupture, the rupture velocity systematically decreases with increasing $\phi$, for both fluid volumes [Fig. \ref{fig:colloid_rupturedata}b]. 

The observation of a constant rupture velocity similar to Culick's law, even at high $\phi$, suggests that the non-Newtonian rheology has a minimal effect on film rupture dynamics. Therefore, it is possible that the rupture of colloidal films may still be modeled as that of a Newtonian fluid. The well-known picture of Newtonian film rupture is that the rim at the rupture boundary collects more and more fluid as it rolls outward. In other words, the information of rupture travels at the same speed as the expanding circular rim, and does not affect  most of the film outside the rupture, except for a small region around the rupture rim, referred to as an aureole~\cite{mcentee1969bursting}. Thus, it is expected that no shear is experienced by the film outside of the aureole. Therefore, our observation that the non-Newtonian rheology has a minimal effect on the rupture, even at high values of $\phi$, is consistent with this mechanism. In addition to introducing non-Newtonian flow behavior, adding colloidal fluids of colloidal particles increases the effective viscosity of the fluid~\cite{mewis2012colloidal, krieger1959mechanism} even in the zero-shear limit. To compare the rupture of colloidal films to its Newtonian counterpart, we used the zero-shear viscosity of suspensions measured via cone-plate rheology. We note that there are several models for predicting suspension viscosity in this limit, and using one of these models instead of experimental viscosity measurements does not alter our conclusions (see SI for more details).

\begin{figure}[!thb]
\centering
\includegraphics[width=0.7\textwidth
]{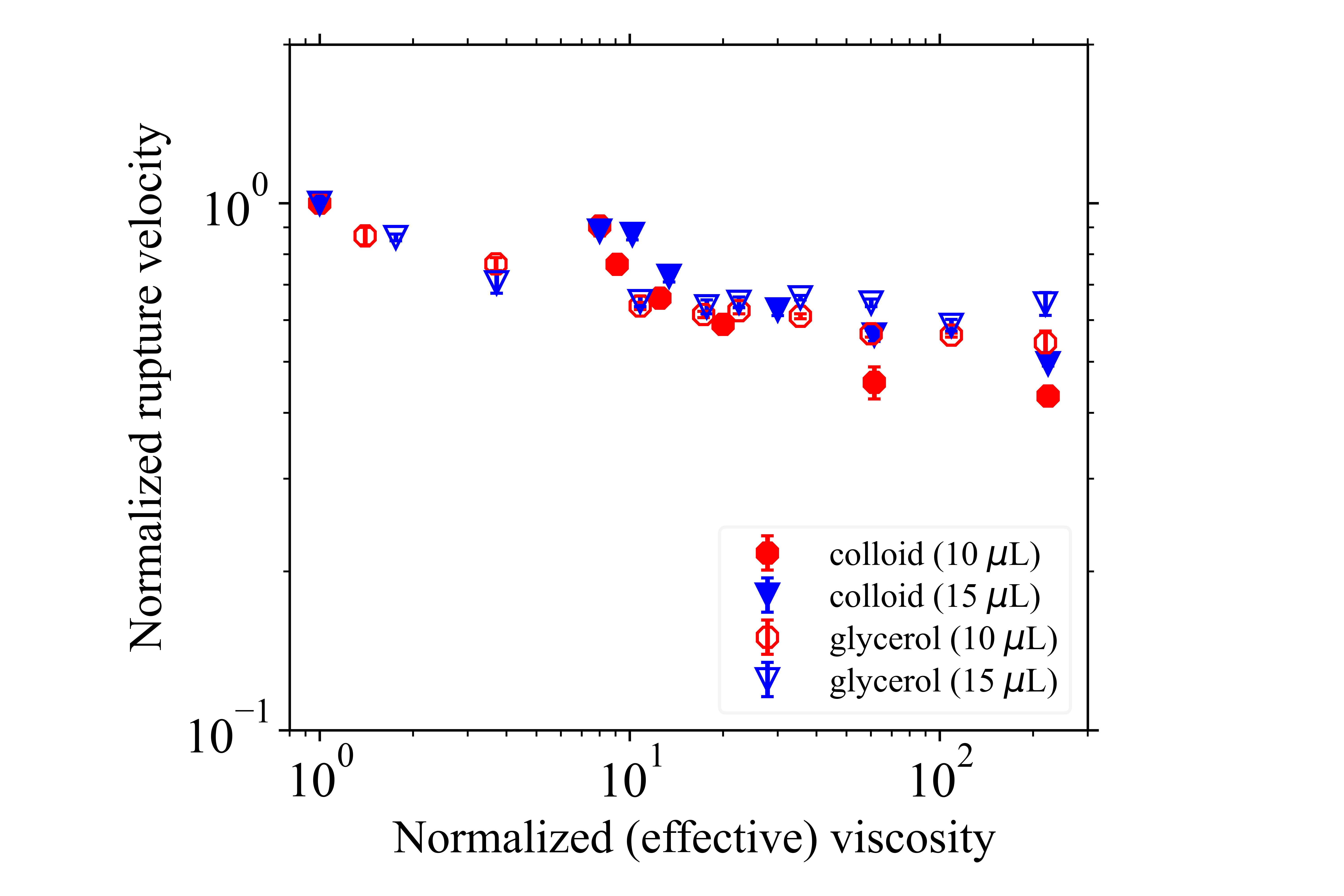}
    \caption[Rupture velocity of colloidal films plotted against the effective viscosity, along with rupture velocity of Newtonian (glycerol-water) films plotted against fluid viscosity, for two fluid volumes.] {Rupture velocity of colloidal films plotted against the effective viscosity, along with Rupture velocity of Newtonian (glycerol-water) films plotted against fluid viscosity, for two fluid volumes.  Both colloidal and Newtonian rupture follow a similar decreasing trend on a log-log plot, indicating that effective viscosity is a useful framework to interpret rupture of colloidal films.   
}
    \label{fig:rupture_viscosity}
\end{figure}

In figure~\ref{fig:rupture_viscosity}, we plotted the film rupture velocity, normalized with respect to $\phi=0$ (soap-water film), against this effective viscosity. Plotted on the same graph is the rupture data of viscous glycerol-water films using the same experimental setup. Both datasets show a similar decreasing trend in normalized rupture velocity. The agreement between these datasets verifies that due to the shear-free conditions in this system, the effective viscosity of the suspension is the dominant parameter that controls film rupture dynamics. We note that while the rupture velocity is constant as predicted by Culick, the magnitude of the rupture velocity decreases with increasing fluid viscosity. Theoretical work has predicted Culick's law to hold even for viscous films after a short initial transient~\cite{savva2009viscous}. Therefore, this decrease in rupture velocity may be attributed to an increase in film thickness for more viscous films, even though we formed our films with the same fluid volume.

\subsection{Film thickness estimation}

Savva et al.~\cite{savva2009viscous} theoretically studied the effect of fluid viscosity on rupture dynamics, and concluded that viscous films rupture at the same speed as Culick velocity, after a brief transient of a timescale that depends on viscosity. For the experimental conditions in this study, this transient is smaller than the temporal resolution of our high-speed camera ($\le 10$ \textmu s). Therefore, according to this model, we should still observe rupture dynamics that follow Culick's law,  and the rupture velocity of constant-thickness films should be independent of fluid viscosity. Our observations of decreasing rupture velocity [Figure ~\ref{fig:colloid_rupturedata}b] with increasing viscosity are clearly in contradiction with this. To explore whether this decrease is attributable to an increase in film thickness, we characterize the film thickness as a function of fluid viscosity. The refractive index mismatch between colloidal particles and the suspending fluid poses a challenge in characterizing film thickness using interferometry or dye measurements. Therefore, we carry out these measurements for Newtonian glycerol-water films.

The side length of the square films formed in our experiments is $L_{film}=25$ mm . Using the two fluid volumes from our experiments and assuming a perfectly flat film, volume conservation allows us to estimate the film thicknesses as:

\begin{equation}\label{eq:thick_10}
    h_{10}=\frac{10\mathrm{\mu L}}{{L_{film}}^2}=16\mathrm{\mu m},
\end{equation}

and 

\begin{equation}\label{eq:thick_15}
    h_{15}=\frac{15\mathrm{\mu L}}{{L_{film}}^2}=24\mathrm{\mu m}.
\end{equation}
We note that these values are upper bounds on the true film thicknesses, as volume conservation gives us a maximum of average film thickness. A more accurate estimate of the true thickness away from film boundary can be made, at least for low-viscosity Newtonian films, using  our rupture velocity data and Culick's law:

\begin{equation} \label{eq:thick_culick_10}
    h_{10, c}= \frac{2\sigma}{\rho {v_{10,c}}^2}=3.2 \mu m,
\end{equation} 

and

\begin{equation}\label{eq:thick_culick_15}
    h_{15, c}= \frac{2\sigma}{\rho {v_{15,c}}^2}=6.2 \mu m,
\end{equation}
where $\sigma$ is the surface tension of 4 mM SDS in water ($45$ mN/m), $\rho$ is the density of water ($1000$ kg/m$^3$), and $v_{10(15),c}$ is the rupture velocity from our experimental data for 10(15) \textmu L films. Thus, the thickness estimated using the Culick velocity is much smaller than the  volume conservation estimate, supporting our hypothesis of a non-trivial film thickness profile.
 
\begin{figure}[!thb]
\centering
\includegraphics[width=\textwidth
]{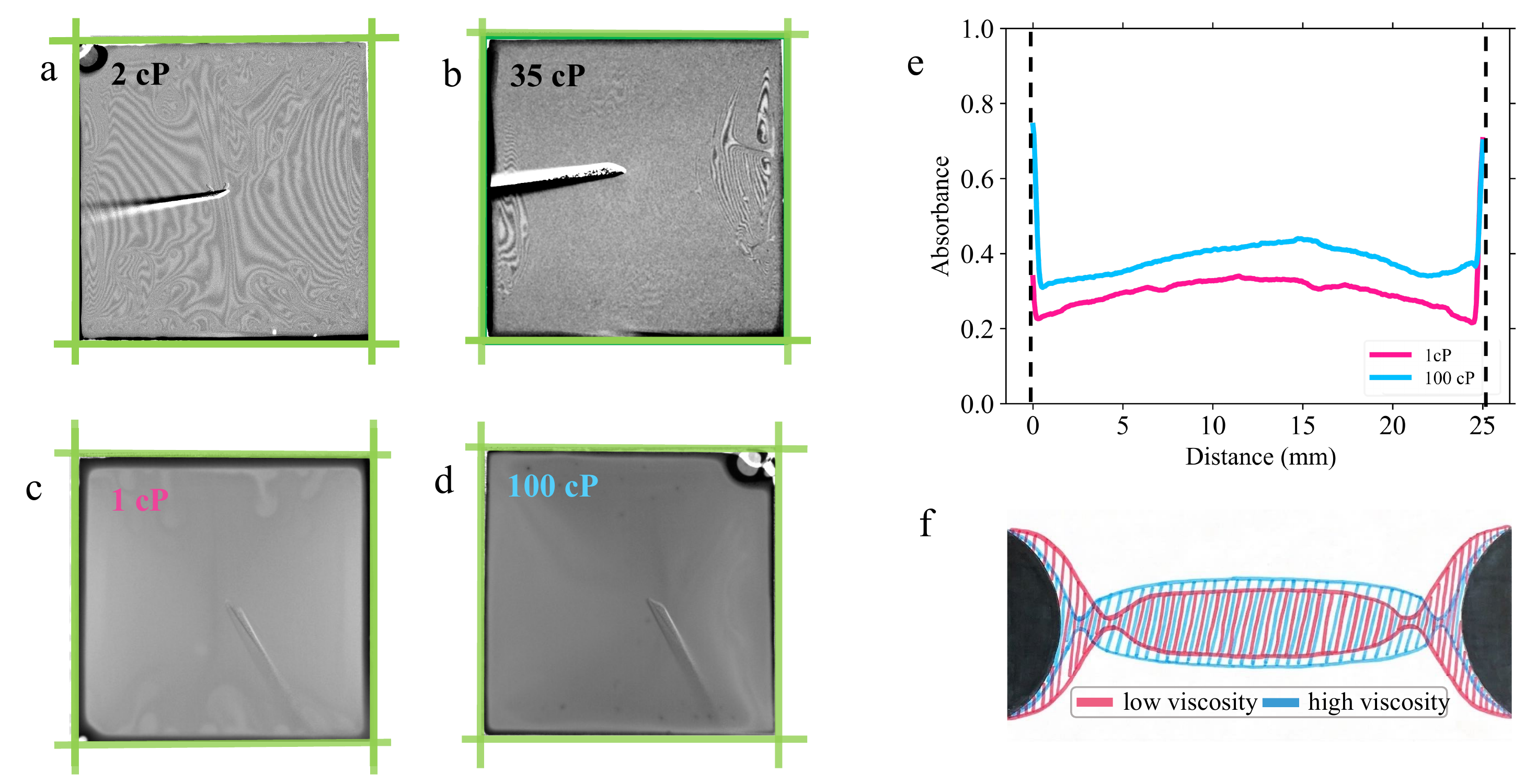}
    \caption{(a) Interference fringes observed under a green filter ($\lambda=$ 510 $\pm$ 10 nm) for a low-viscosity film (2 cP), and  (b) a higher-viscosity film (35 cP). The change in the interference pattern indicates that the thickness profile changes with fluid viscosity. (c,d) 15 \textmu L Newtonian films of viscosities \textbf{c} 1 cP and  \textbf{d} 100 cP, containing Brilliant Blue dye. The films transmit different amounts of light, indicating a different film thickness. (e) Absorbance of both the films plotted along a line going through the center of the film. Higher absorbance indicates higher film thickness, thus the film thickness profile is proportional to the absorbance profile. (f) Schematic showing the hypothesized variation in thickness profile with increasing viscosity. The pink shaded area indicates a lower-viscosity film, and the blue indicates a higher viscosity film made of the same fluid volume. The length of the meniscus near the film edge is smaller for a more viscous film, and so is the location where film thickness is minimum. Away from the boundary, both films assume a flat, practically constant-thickness shape, with the more viscous fluid forming a thicker film. Features are exaggerated for clarity.
}
    \label{fig:fringes}
\end{figure}

 Interferometry is commonly used to characterize the thickness profile of thin fluid films~\cite{He_2020}. Figure~\ref{fig:fringes} shows images of films made from the same fluid volume (15 \textmu L), but with fluids of different viscosities under a narrow-width bandpass filter ($\lambda=530 \pm 10$ nm). The fluid viscosity of the film imaged in Figure~\ref{fig:fringes}a is $\eta=2$ cP, and that showed in Figure~\ref{fig:fringes}b is $\eta=35$ cP. While the $2$ cP film shows distinctive interference fringes, fringes are largely absent for the $35$ cP film, except near the film boundary. The absence of fringes in Figure~\ref{fig:fringes}b indicate that the film thickness away from the boundary is larger than the coherence length for the bandpass filter used. The coherence length can be calculated as:

\begin{equation}
    l_{c}=C\frac{\lambda^2}{n\Delta \lambda}, 
\end{equation}
where $n$ is the refractive index of the medium (calculated for every film composition using the refractive index of 1.33 for water and 1.45 for glycerol) and $C=0.44$~\cite{Akcay_coherence_length};
For $\lambda=530$ nm and $\Delta \lambda= 10$ nm, we obtain $l_{c}=9.3$ \textmu m. Thus, we infer that the higher-viscosity (35 cP) film is thicker than the low-viscosity (2 cP) film, despite being made of the same volume of fluid. As thicker films rupture at lower speeds according to Culick's law, this is also consistent with the lower rupture speed we observed for viscous films. Additionally, in Figure~\ref{fig:fringes}b, we observe an interference pattern near the film edge, although fringes are absent near the middle. This suggests that the film has a non-monotonic thickness profile. To characterize this film thickness profile in more detail, we collect dye absorbance data for films of varying viscosity.

Figures ~\ref{fig:fringes}c, d show images of fluid films of viscosities $1$ cP and $100$ cP respectively, containing 10 g/L Brilliant Blue dye and imaged under white light. For dyed fluids, the absorbance gives the fraction of light absorbed by the sample, and is calculated as:

\begin{center}
    Absorbance $=1-$ normalized intensity of transmitted light.
\end{center} 

Figure~\ref{fig:fringes}e shows a plot of absorbance for both films, across a line drawn through the center of the film. In order to minimize noise, the profile averaged over a 10-pixel wide box around the line is plotted. As a thicker sample absorbs more light, the absorbance is proportional to the film thickness. Thus, the absorbance profile qualitatively captures the features of the film thickness profile. The more viscous film (blue line) shows a higher absorbance near the film center, further evidence that more viscous films lead to thicker films despite being made from the same fluid volume.

Two other features of the absorbance profile are particularly informative. First, the absorbance increases rapidly near the film boundary, indicating a drastic increase in film thickness. This can also be observed in the Figure~\ref{fig:fringes}c and d, in the form of a darker region near the wire frame. Near the boundary, the film thickness must approach the thickness of the wire frame ($500$ \textmu m),  causing this thicker meniscus. The second interesting feature is that the absorbance reaches its minimum as we move towards the center of the film from the meniscus, well before the film assumes a relatively flat shape. The thinnest part of the film is therefore located next to the meniscus near the boundary, and not near the center of the film. Thus, the absorbance of dyed films verifies that fluid viscosity directly controls the film thickness profile. Quantifying film thickness with dye absorbance data is difficult, as making calibration samples as thin as a few microns is experimentally challenging. Nevertheless, the qualitative features of the absorbance profile allow us to construct a sketch of the film thickness profile. 

\begin{figure}[!htb]
\centering
\includegraphics[width=0.9\textwidth
]{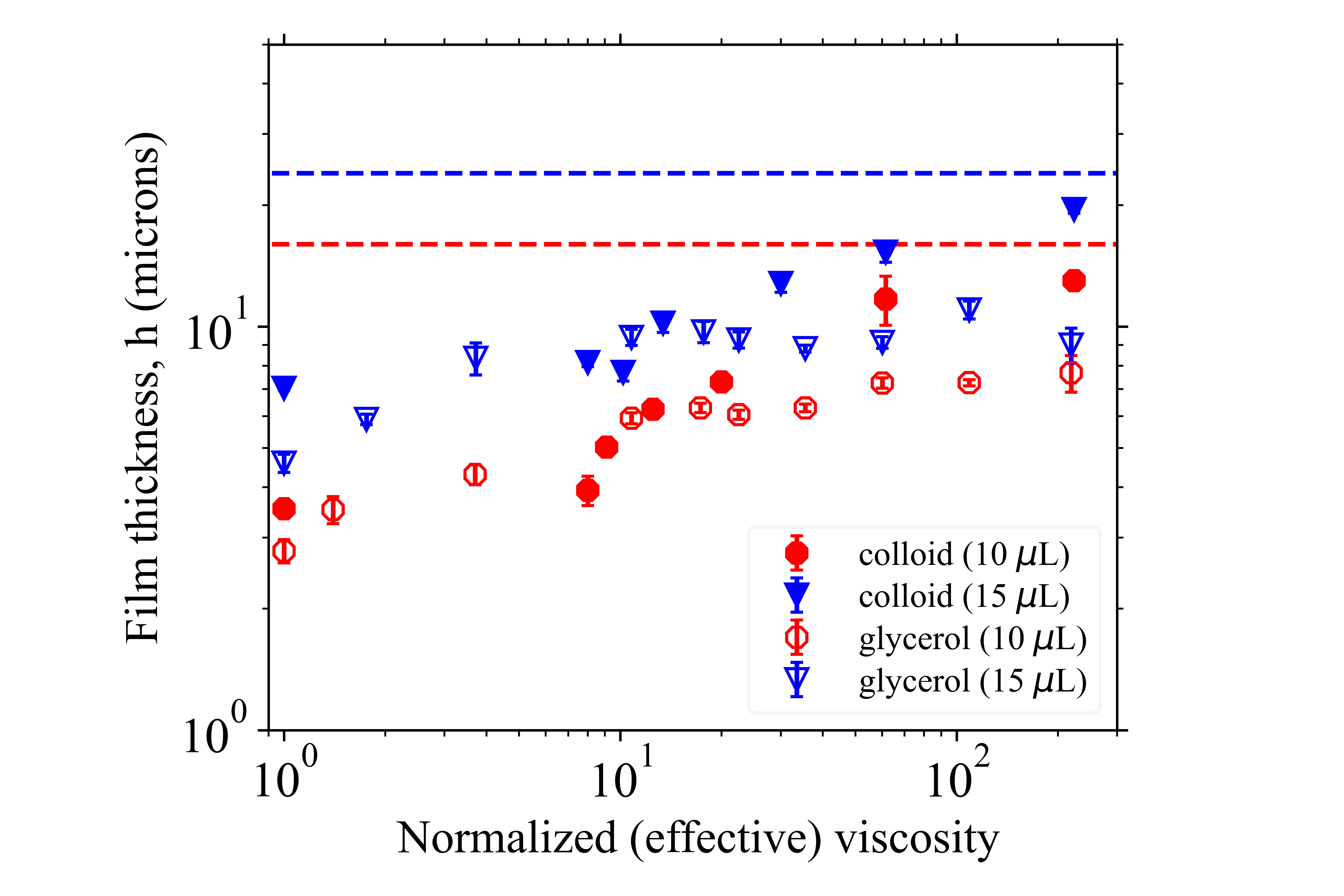}
    \caption [Film thickness near the film center, calculated using Culick's law, plotted for both Newtonian and colloidal films.]{Film thickness, calculated using Culick's law, plotted for both Newtonian and colloidal films. For both viscous and colloidal fluids, the film thickness increases with increasing fluid viscosity, consistent with the interferometry and dye absorbance observations. 
}
    \label{fig:h_vs_eta}
\end{figure}

Figure~\ref{fig:fringes}f shows a sketch of the thickness profile inferred from our interferometry and dye absorbance measurements, as seen from the side. This is a schematic, and features are exaggerated for clarity. Films have a large thickness (of the order of the wire frame thickness) near the boundary, which dramatically decreases as we move away from the boundary. The thickness reaches a minimum before rising again, and achieves a relatively flat profile near the center. The more viscous film (blue shaded region) is thicker in the middle than the low-viscosity film (pink shaded region), leading to a slower film rupture [Figure~\ref{fig:rupture_viscosity}]. This schematic also demonstrates how the same fluid volume can result in films of different thicknesses. This occurs, because the meniscus near the film edge can be made of different fluid volumes, as evidenced from the crossover point between the blue and pink profiles. Thus, the area contained by the two schematic profiles is the same, corresponding to films containing the same fluid volume. This result is consistent with the variation of film thickness with viscosity observed in related systems. Our system shares similarities with a film clinging to a fiber pulled slowly out a reservoir of fluid. LLD theory predicts that this film thickness grows with increasing viscosity~\cite{deGennes2004capillarity,MYSELS1962FrankelExp}. As the glycerol-water films considered in our experiments have viscosity values that are within the applicable range of the LLD approximation, the increasing thickness trend is qualitatively consistent with the LLD prediction in a related system.

Thus, our characterization of the film thickness profile indicates that when a constant volume of fluid is used to make horizontal flat films, the film thickness profile --- and in turn the film thickness away from the boundary --- is directly set by the fluid viscosity. Consequently, the film thickness affects the rupture velocity. Therefore, our results are consistent with the prediction by Savva and Bush~\cite{savva2009viscous} that Culick's law is asymptotically applicable even for the rupture of viscous films. Using equation \ref{eq:culick}, we convert the rupture velocity into film thickness: $h=\frac{2\sigma}{\rho v^2}$. Figure~\ref{fig:h_vs_eta} shows the thickness thus calculated plotted against the fluid viscosity (for colloidal data, effective fluid viscosity), showing an increasing trend in film thickness with viscosity. The dotted lines indicate the upper-bound film thickness estimated from volume conservation, Equations~\ref{eq:thick_10} and \ref{eq:thick_15}.

\subsection{Instabilities in spontaneously rupturing colloidal films}

\begin{figure}[!htb]
\centering
\includegraphics[width=\textwidth
]{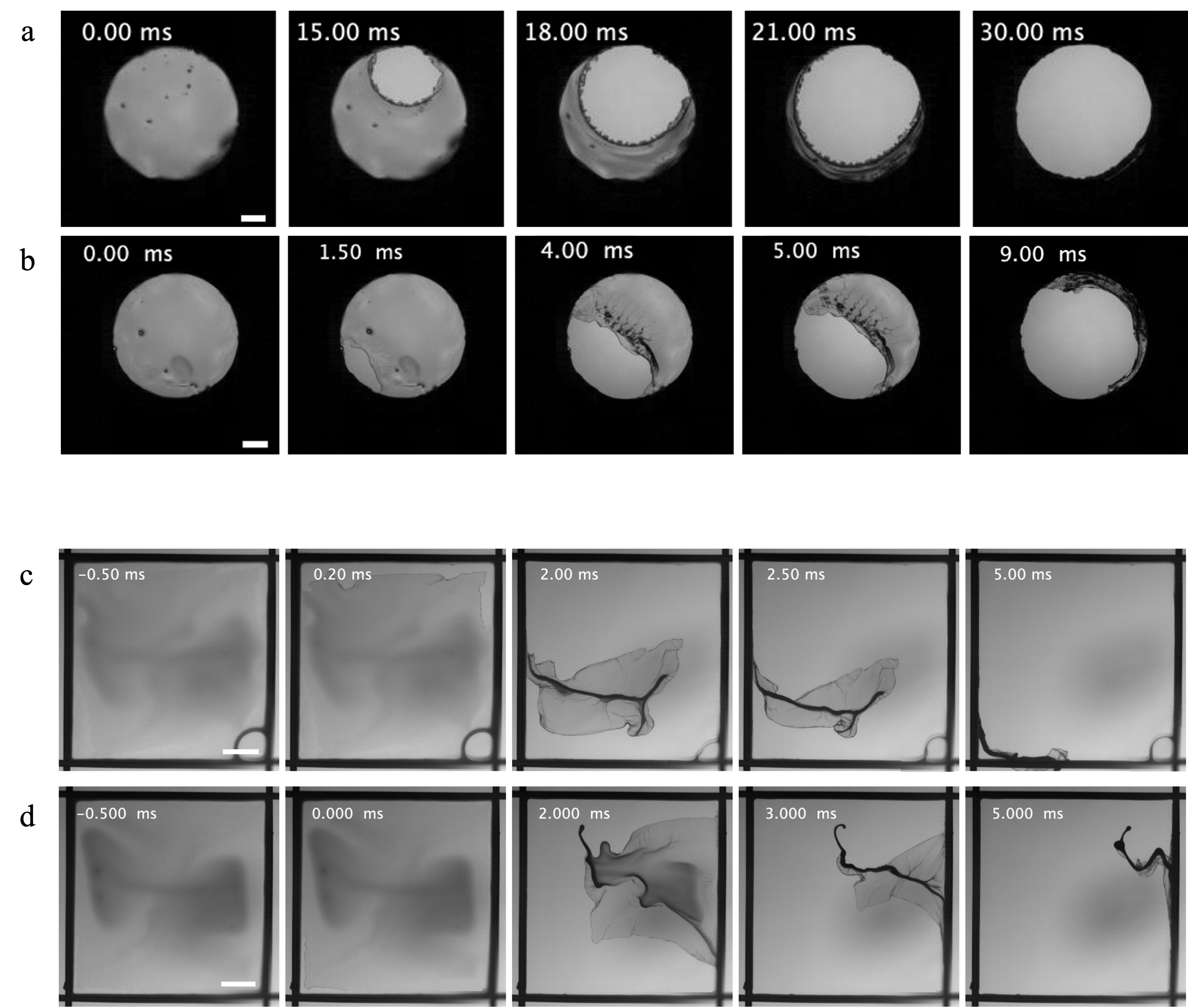}
    \caption [Instabilities develop in spontaneously rupturing colloidal films.]{(a,b) Instabilities develop in spontaneously rupturing colloidal films at low humidity.  The film composition is $\phi=0.35$ colloidal particles, where the  suspending fluid is 2mM SDS in 30\% glycerol and 70\% water. In uncontrolled humidity, instabilities are not reproducible (see also SI videos 2,3~\cite{filmSI}). (c,d) Instabilities also develop in spontaneously rupturing colloidal films at high humidity. The film was made of fluid with $\phi=0.40$ colloids in 4mM SDS and water (see also SI video 4~\cite{filmSI}). In a humidity-controlled environment, instabilities are much more reproducible. Scale bars are 2 mm. 
}
    \label{fig:instabilities}
\end{figure}

In contrast to films manually ruptured away from the boundary, when constant-volume colloidal films were allowed to rupture spontaneously, we observed instabilities developing near the rupture boundary and propagating into the intact film during rupture [Figure ~\ref{fig:instabilities}, SI videos 2,3,4~\cite{filmSI}]. Spontaneous rupture always happened near the film edge; the instabilities were almost nonexistent when similarly prepared films were manually ruptured away from the boundary. Our qualitative characterization of the film thickness profile [Figures~\ref{fig:fringes}e,f] suggests that the film is thinnest near the edge, between the thicker meniscus near the wire frame and the approximately flat film near the center [Figure \ref{fig:fringes}f]. Thus, our observation that spontaneous rupture always occurs near the edge is consistent with the film thickness profile. 

For the experiments where we observed instabilities, we formed colloidal fluid films of about 12 mm in size with 2 \textmu L of fluid, and recorded the film until it spontaneously ruptured. We observed instabilities in experiments where humidity was not controlled for [Figure \ref{fig:instabilities}a], and also in controlled humidity experiments inside the humidity chamber [Figure \ref{fig:instabilities}b, SI video 4~\cite{filmSI}]. Figures \ref{fig:instabilities}a and b show $\phi=0.35$ colloidal films, where the  suspending fluid is 2mM SDS in 30\% glycerol and 70\% water, rupturing at ambient humidity (RH $\sim$ 30\%). We note that these experiments were done using an earlier version of the setup where films were made using an expandable camera iris, leading to a quasi-circular film geometry. Scale bars in all panels are 5 mm. In Figure~\ref{fig:instabilities}a (see also SI video 2~\cite{filmSI}), we observe that the rupture edge has protruding jagged structures. Additionally, the intact film develops wrinkles that crowd together as the rupture proceeds. These wrinkles are parallel to the moving rupture rim. Figure~\ref{fig:instabilities}b see also SI video 3~\cite{filmSI}), on the other hand, shows instabilities in the transverse direction to the rupture rim. These instabilities appear near the rim and propagate outward ahead of the rupture. We note that both these films were made of the same fluid, and yet they show drastically different patterns during rupture. Additionally, the rupture time of the two films differs by a factor of 3. This lack of reproducibility may potentially be attributed to large fluctuations in the ambient humidity, thus indicating that evaporation may have a significant effect on film dynamics and the progression of instabilities.

Figures \ref{fig:instabilities}c and d show 12 mm square films with $\phi=0.40$ in 4 mM SDS and water, rupturing under controlled humidity (RH= 80\%). Once again, the rupture begins near the edge, where the film is thinnest. We observe instabilities originating from the rupture rim and propagating throughout the film. A thicker (darker) structure forms in the intact film, and then the rest of the film appears to wrinkle like a fabric around this thicker filament. After rupture, we observe the filament to flow like a fluid, indicating that evaporation has not dried out the film (see also SI video 4~\cite{filmSI}). In these humidity-controlled experiments, both the qualitative rupture behaviour and the rupture time were reproducible over multiple trials.

For both humidity-controlled and uncontrolled experiments, instabilities were predominantly observed for spontaneously rupturing films. This behaviour has several striking features. In the beginning of rupture, the rupture boundary is jagged as opposed to the smooth rupture boundaries for manually ruptured films. While the rupture rim increases in size as it collects more fluid for manually ruptured films, we do not detect such widening of the rim in case of films where instabilities are observed. The rupture in this case looks qualitatively different: the manually induced rupture rim rolls outward smoothly, while the spontaneous rupture with instabilities is similar to an elastic sheet de-pinning from the wire frame. The instabilities are reminiscent of folds or wrinkles on a fabric. Even for controlled humidity, the significant variation in the shape of the rupture front and the location of instabilities makes quantifying their dynamics a challenge. Furthermore, the structures observed in ambient environment experiments varied between multiple trials, indicating a possibility of extreme sensitivity to initial conditions. This might be a result of the significant amount of stochasticity contributing to film rupture by fluctuating humidity and ambient impurities.

We observe these instabilities only for a specific set of conditions: They occurred at high values of colloidal $\phi$, and for relatively thinner films, i.e.\ films made of a smaller volume of fluid. Additionally, spontaneously rupturing films developed instabilities that spanned the whole size of the film. We hypothesize that these structures develop when the colloidal size (660 nm) competes with the film thickness (a few microns). For films to spontaneously rupture, we waited for rupture after forming the film. Over this waiting time, the film could have thinned due to drainage. Additionally, fluid films are more likely to spontaneously rupture at a location where they are the thinnest~\cite{neel_villermaux_2018_thickfilms}. Our flat films are thinnest close to their boundary [Figure \ref{fig:fringes} e,f]; this is a possibly why instabilities develop in spontaneously rupturing films.  Another study has reported fold-like instabilities even in collapsing Newtonian films, when rupture was initiated at the film boundary~\cite{habibi2021foldsGlycerol}. The origin of these folds was attributed to the geometric singularity caused by the sharp corners of the frame, as folds were absent in a smoother frame geometry. Thus, the initiation of the rupture near the square frame in our spontaneous rupture experiments may also contribute to the occurrence of these instabilities. 

For controlled humidity experiments, although the nontrivial geometry of the rupturing film made it challenging to measure the rupture speed, we can estimate the rupture speed using the film side length (12 mm) and time of rupture (7 ms), $v_{rupture}=1.7$ m/s. If the rupture were to follow Culick's law, this rupture speed would correspond to a 22 \textmu m film. This is clearly unphysical, as this thickness is above the upper bound of thickness estimated using volume conservation, 13 \textmu m. Therefore, instabilities significantly slow down rupture. In other systems, such a slowing down has been attributed to film elasticity~\cite{petit2015holes}. A deeper investigation into the cause of these modified film dynamics would provide more information about the microscopic dynamics in colloidal films.

\section{Conclusion}

Here, we report the rupture velocity data of flat colloidal films in the range of volume fractions $0\le \phi\le 0.47$, and of glycerol-water films with viscosities 1-235 cP.  We use a custom film stretcher to make films of two constant fluid volumes, 10 \textmu L and 15 \textmu L. We observe constant rupture velocity even at high colloidal volume fractions which, when plotted against the effective viscosity of the suspensions, agrees well with that of Newtonian glycerol-water films in the same viscosity range. Therefore, even at high $\phi$ where highly non-Newtonian flow behaviour is apparent from bulk rheology, the effective suspension viscosity is sufficient to capture the dynamics of colloidal film rupture. As the well-accepted picture of Newtonian film rupture is that the rupture rolls outward collecting fluid, high shear must not be present outside of the film aureole during rupture, despite the short timescale of rupture. Our results demonstrate that rupturing particulate films can be effectively modeled as viscous fluids.  

For both colloidal and Newtonian fluids, we observed the rupture velocity of constant-volume films decreases with increasing fluid (effective) viscosity. Our characterization of the film thickness profile via interference imaging and dye absorption shows that higher viscosity films lead to higher thickness near the film center, as viscosity modifies the overall thickness profile. This increase is consistent with the predictions of LLD theory in the visco-capillary regime~\cite{deGennes2004capillarity, mysels1959soap}.

Although Culick's derivation of rupture velocity neglected viscous effects, Culick's law has been predicted to asymptotically hold for high-viscosity fluids~\cite{savva2009viscous}. Viscosity is only expected to introduce a transient at the initial stage of rupture, on a timescale $t_{vis}=\frac{\eta H}{2\sigma}$. For our experiments, $0.1$ \textmu s $\le \tau_{vis} \le  20$ \textmu s, which cannot be observed at the temporal resolution of our experiments ($\sim 10$ \textmu s). Thus, the change in the thickness profile is solely responsible for the slower rupture of more viscous films. A surprising feature of the horizontal film thickness profile is that the film reaches a minimum thickness near the thicker meniscus by the wire frame, away from the film center. The reason for this highly non-trivial film shape should be studied carefully in the future. 

When we allowed flat colloidal films at $\phi\ge0.40$ to rupture spontaneously, we observed exotic instabilities originating at the rupture rim and propagating through the film surface. Films always spontaneously rupture near the edge, as films have a thickness minimum near the edge that increase the probability of rupture. Both the pattern of instabilities and the rupture time were highly reproducible under controlled and high humidity, while they were much more stochastic when humidity was not controlled for. We hypothesize that these structures occur when the colloidal size is comparable to the film thickness. Other studies have reported instabilities in surfactant films of varying thicknesses above the critical micellar concentration, which have been attributed to the rigidity imparted to the film~\cite{petit2015holes} and to micelles forming mesa-like structures~\cite{zhangSharma2015domain,VivekSharma_mesa_2021}. Some work has also focused on particulate rafts with $> 100$ \textmu m particles, so that the raft is made of small capillary bridges between these particles~\cite{timounay2015particulate}. However, there are no other works that have considered colloidal-size particles in films that are less than an order of magnitude thicker than the particle, to the best of our knowledge. Film rupture in this parameter regime is a very convenient way to study the physics of suspensions in two dimensions. Rupturing films are reminiscent of wrinkling fabrics, and instabilities seem to slow down film rupture significantly as compared to the Culick prediction. The characterization of this state of the film before it re-fluidizes after rupture would be a fascinating avenue of future exploration.  

Our experiments on colloidal film rupture demonstrate that our understanding of Newtonian film rupture can be extended to the rupture dynamics of non-Newtonian films. Despite colloidal film dynamics being surprisingly Newtonian when the films are significantly thicker than particle size, we observed exotic structures in spontaneously rupturing films when the film thickness and particle size were comparable. Further theoretical work aimed towards understanding these discrete effects on the particle scale would greatly enhance our understanding of changes in colloidal microstructure under dynamical conditions in two dimensions. We anticipate that our work will guide future applications that necessitate a controlled use of fluid films in a variety of applications, such as surface coatings and petrochemical recovery.

\section*{Acknowledgements}
The authors are thankful to Vivek Sharma for useful discussions and insights. 

\newpage

\section*{Supplemental Information}

\begin{figure}[!thb]
\centering
\includegraphics[width=0.5\textwidth
]{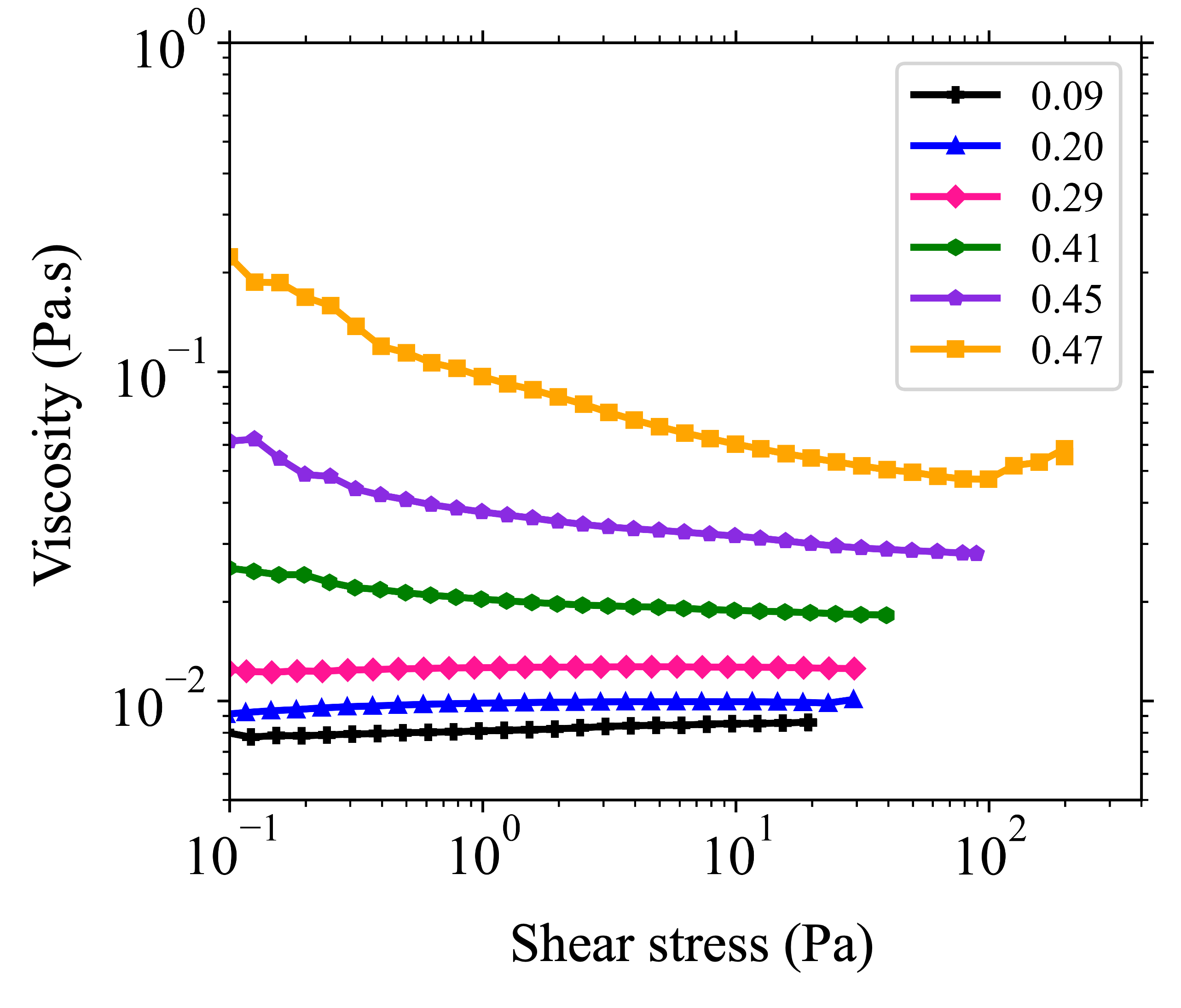}
    \caption*{FIG.\ S1. Rheology of 830 nm sphere suspensions for $0.09\le\phi\le0.47$. We use the value of the zero-shear rheology (first point in each dataset) to represent rupture data in terms of the effective suspension viscosity. 
}
    \label{fig:rheo}
\end{figure}

In order to represent rupture data in terms of the effective suspension viscosity, we used the zero-shear viscosity of our colloidal suspensions from rheology data. Figure S1 shows the viscosity of 830 nm colloidal suspensions plotted against shear stress, for the range of $\phi$ relevant in our rupture experiments. This data was collected using cone-plate rheology, with a 1° cone angle. As the film rupture process is shear-free, we used the first point on the rheology curve as the effective viscosity value. Although the spheres used in our experiments were 660 nm (smaller than the ones used for rheology, 830 nm), such a small variation in colloidal size does not significantly alter the zero-shear viscosity of the suspension~\cite{mewis2012colloidal}. 
\begin{figure}[!thb]
\centering
\includegraphics[width=0.5\textwidth
]{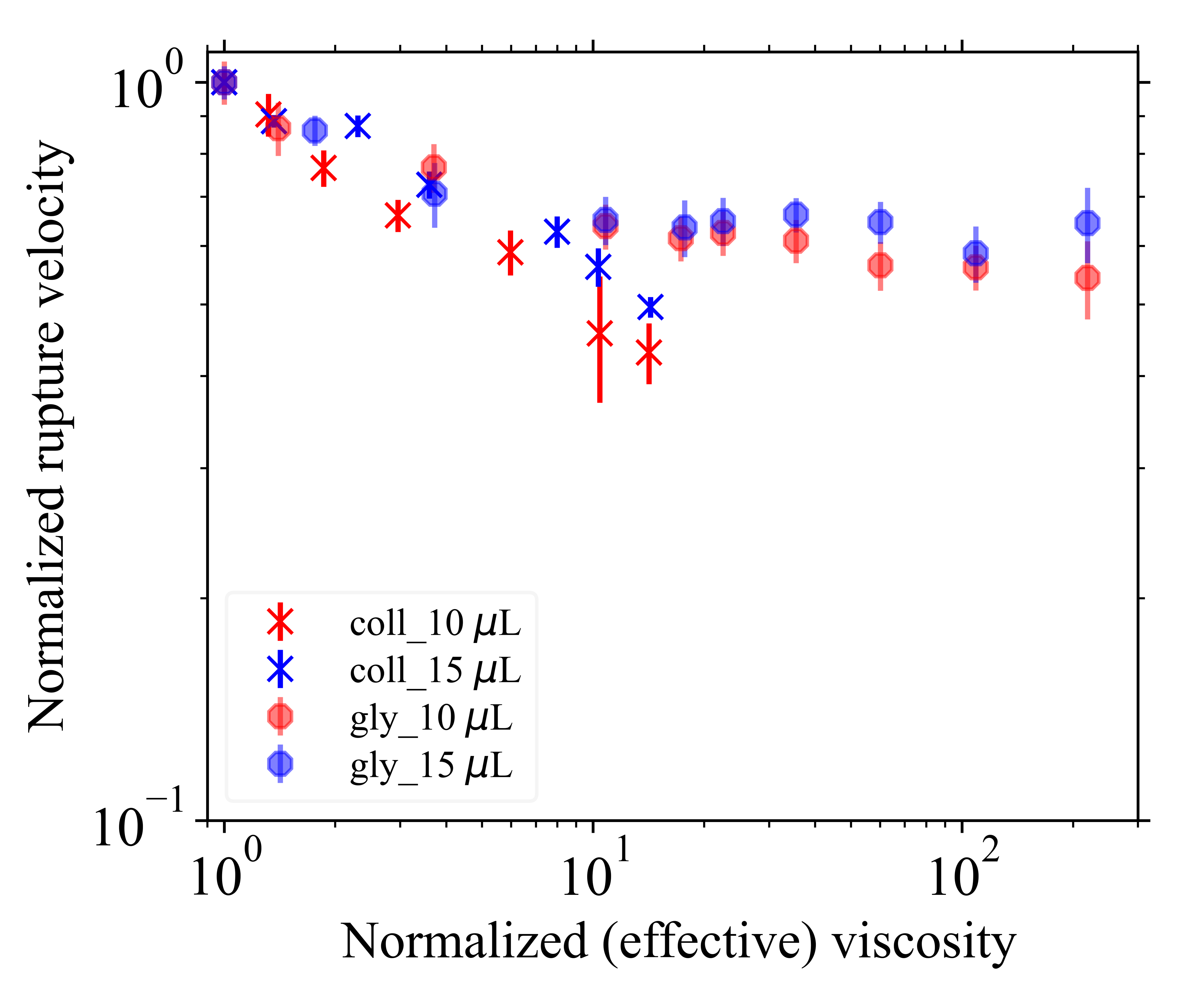}
    \caption*{FIG.\ S2. Film thickness, calculated using Culick's law, plotted for both Newtonian and colloidal films against fluid viscosity normalzied with the viscosity of water. For both viscous and colloidal fluids, the film thickness increases with increasing fluid viscosity, consistent with the interferometry and dye absorbance observations. 
}
    \label{fig:rupture_visc_KD}
\end{figure}
Figure S2 shows the normalized rupture velocity plotted against this K-D effective viscosity~\cite{krieger1959mechanism}. Although the overlap between colloidal and viscous data is altered with this choice of effective viscosity, plotting in terms of this quantity does not change our conclusion that the effective viscosity framework is relevant for interpreting colloidal rupture dynamics.
We note that estimating the effective viscosity via the Krieger-Dougherty model vastly underestimates the zero-shear viscosity in case of our suspensions. Over $0.00\le\phi\le0.47$, the K-D model predicts viscosity values that range from 1 cP to 11 cP, while rheology data suggests the viscosity ranges from 1 cP to over 200 cP. The colloidal suspensions considered in our experiments are charge stabilized, and not hard spheres. Thus, the hydrodynamic size of the particles is larger than the actual particle size due to electrostatic repulsion. We believe that the K-D model underestimates the zero-shear viscosity because of this difference in the particle size, and in turn the effective volume fraction $\phi$ of the suspension. Therefore, we believe that using the experimentally measured zero-shear viscosity of the suspensions is the most faithful representation of the suspension viscosity.

\bibliographystyle{apsrev4-2}
\bibliography{film_bib}

\end{document}